\documentclass[prl,twocolumn]{revtex4}
\usepackage{graphicx}

\begin{document}

\def\K{{\bf{K}}}
\def\Q{{\bf{Q}}}
\def\Gbar{\bar{G}}
\def\tk{\tilde{\bf{k}}}
\def\k{{\bf{k}}}

\title{Suppression of superconductivity in the Hubbard model by buckling and breathing phonons}
\author{Alexandru Macridin$^{1}$, Brian Moritz$^{1,2,3,4}$, M.\ Jarrell$^{1}$, 
Thomas Maier$^{5}$}
\address{
$^{1}$ University of Cincinnati, Cincinnati, Ohio, 45221, USA\\
$^{2}$ University of Waterloo, Waterloo, Ontario, N2L 3G1, Canada \\
$^{3}$ Pacific Institute of Theoretical Physics, University of British Columbia, British Columbia, V6T 1Z1, Canada \\
$^{4}$ University of North Dakota, Grand Forks, North Dakota, 58202, USA\\
$^{5}$ Oak Ridge National Laboratory, Oak Ridge, Tennessee, 37831, USA}

\date{\today}

\begin{abstract}

We study the effect of buckling and breathing phonons, relevant for cuprate superconductors, 
on the $d$-wave superconductivity in the two-dimensional Hubbard model by employing dynamical 
cluster Monte Carlo calculations. 
The interplay of electronic 
correlations and the electron-phonon interaction  produces two competing 
effects, an enhancement of the effective $d$-wave pairing interaction, which favors 
$d$-wave superconductivity, and a strong renormalization  of the single-particle 
propagator, which suppress superconductivity. Due to the later effect we 
find that buckling and breathing phonons  suppress the superconductivity
in the region of parameter space relevant for cuprate superconductors.
\end{abstract}
\maketitle


{\em{Introduction.}} 
Many experiments, including Raman~\cite{raman}, neutron scattering~\cite{neutron},
and photoemission~\cite{lanzara,gweon} find clear evidence of electron-phonon (EP) 
interaction in high $T_c$ materials.  While  it is widely accepted that strong 
electronic correlations play a fundamental role in the mechanism of high 
temperature superconductivity, the role of EP interaction is under hot debate.

The phonons believed to be the most relevant to the cuprates are the buckling 
(BC) and the breathing (BR) modes~\cite{neutron,raman,lanzara,rosch,OKA,cuk}, 
defined, respectively, by the out-of-plane and in-plane displacement of the 
oxygen ions.  Previous investigations suggest that the symmetry of these
phonon modes is significant.   There are arguments suggesting that coupling 
to the out-of-plane oxygen BC or the local Holstein (H) phonons, enhance the 
$d$-wave pairing interaction~\cite{bulut,t_sakai_97,huang:qmc,piekarz,honerkamp},  
whereas, the BR mode has been found to suppress the interaction~\cite{bulut,t_sakai_97,piekarz}.
It also was shown that the  electronic correlations in the presence of EP 
coupling strongly enhance polaron formation~\cite{j_zhong_92,p_prolovsek_06,macridin:hh} 
and renormalize the quasiparticle (QP) weight, an effect which suppresses  
superconductivity. 

In this letter we employ the dynamical cluster approximation 
(DCA)~\cite{hettler:dca,maier:rev} to study 
two models relevant for cuprates, the Hubbard model with  BC and  BR phonons.
We also extend the work on superconductivity in
the Hubbard-Holstein model~\cite{macridin:hh}. DCA has proved to be one of the few techniques 
able to capture the superconducting properties
of strongly correlated systems~\cite{maier:rev}.
We investigate the role of phonons on superconductivity in the region of 
small and intermediate doping where the antiferromagnetic (AF)
correlations are strong. 
In agreement with previous investigations, we find 
enhancement of the $d$-wave pairing interaction by H~\cite{macridin:hh,huang:qmc} and 
BC~\cite{bulut,t_sakai_97,piekarz,honerkamp} phonons.
Moreover we find an enhancement of the $d$-wave pairing interaction even 
for BR phonons contrary to some  predictions~\cite{bulut,t_sakai_97,piekarz}.
However, despite the enhancement of the $d$-wave pairing, we find a 
suppression of the superconducting $T_c$ for all three modes, due to the 
strong renormalization of the electronic single-particle propagator. 

{\em{Formalism.}}
The Hamiltonian for each model, Hubbard-Holstein (HH), 
Hubbard-Buckling (HBC) and Hubbard-Breathing (HBR), can be written as
\begin{eqnarray}
H=H_U+H_{ph}+H_{ep}
\label{eq:ham}
\end{eqnarray}
\noindent where
\begin{eqnarray}
H_U=-t \sum_{\langle ij\rangle\sigma}\ \left(c^\dagger_{i\sigma} c_{j\sigma} +
c^\dagger_{j\sigma} c_{i\sigma}\right) + U \sum_i n_{i\uparrow} n_{i\downarrow}
\label{eq:hamu}
\end{eqnarray}
\noindent is the Hubbard part  with 
nearest-neighbor hopping $t$ and on-site repulsion $U$. 
For H phonons, described as a set of independent oscillators 
at every site $i$ which couple locally to the electronic density,
\begin{eqnarray}
H_{ph}^{H}+H_{ep}^{H}=\sum_i \frac{p_i^2}{2M} + \frac{1}{2} M \omega_0^2 u_i^2 + g  n_i u_i,
\label{eq:hamphh}
\end{eqnarray}
\noindent where $\{u_{i},\,p_{i}\}$ are canonical conjugate coordinates for 
each oscillator with characteristic frequency $\omega_{0}$. The BC (BR) 
phonons are independent out-of-plane (in-plane) oscillators on each bond
$i+\hat{\gamma}/2$, with $\hat{\gamma}=\hat{x},\hat{y}$, such that
\begin{eqnarray}
\label{eq:hamphbb}
& H_{ph}^{BC,BR}+H_{ep}^{BC,BR}= \\ \nonumber
& =\sum_{i,\gamma} \frac{p_{i+\hat{\gamma}/2}^2}{2M} 
  + \frac{1}{2} M \omega_0^2u_{i+\hat{\gamma}/2}^2 
  + g (n_i \pm n_{i+\hat{\gamma}}) u_{i+\hat{\gamma}/2}
\end{eqnarray}
\noindent where the last term of Eq.~\ref{eq:hamphbb} with plus (minus) 
describes the coupling to the  BC (BR) phonons.  The dimensionless EP coupling, 
defined as the ratio of the single-electron lattice deformation energy to half 
of the electronic bandwidth  $W/2=4t$, is  $\lambda^{H}=2 g^2/(2M\omega_0^2 W)$ 
and $\lambda^{BC,BR}=4 \times 2 g^2/(2M\omega_0^2 W)$~\footnote
{This definition of $\lambda$, common in polaron studies and used 
also in Ref.~\cite{gunnarsson} for cuprates, does not always coincide with the Migdal-Eliashberg 
definition~\cite{sezlak}.}, with an extra factor of 
four in $\lambda^{BC,BR}$  due to local coordination.
Note that in general the coupling to the BC
mode implies both modulation of orbital energy and  hopping integrals~\cite{OKA}.
However, due to the technical difficulties  associated with
off diagonal EP coupling, we consider here only the former effect
as was done previously in Ref.~\cite{bulut,t_sakai_97}.

To study the Hamiltonian (\ref{eq:ham}) we employ the 
DCA, a cluster mean-field theory which for a two
dimensional system  maps the original lattice model onto a
periodic cluster of size $N_c=L_c^2$ embedded in a self-consistent
host. Correlations up to a range $L_c$ are treated explicitly, while
those at longer length scales are described at the mean-field level.
With increasing cluster size, the DCA systematically interpolates
between the single-site dynamical mean field result and the exact result, 
while remaining in the thermodynamic limit. Cluster mean field calculations 
on the simple Hubbard model successfully reproduce
many of the features of the cuprates, including a 
Mott gap and strong AF correlations, 
d-wave superconductivity and pseudogap behavior~\cite{maier:rev}.

We solve the cluster problem using a quantum Monte Carlo (QMC)
algorithm~\cite{jarrell:dca} modified to perform the sum over both the
discrete field used to decouple the Hubbard repulsion, as well as the
phonon field $u$.  The space of configurations of the latter field is
significantly larger than the former.  In part, this is offset by the
strong correlations of the phonon field in Matsubara time.  Therefore,
in the QMC Markov process, correlated changes, in which adjacent
phonon fields in time are moved together, are mixed in with local
moves to reduce the autocorrelation time.  Nevertheless, the present
code including the effect of phonons requires significantly more CPU 
time than required for the Hubbard model.  Thus, 
most of the  present calculations are restricted to clusters of size $N_c=4$.
Nevertheless we check the robustness of our conclusions by employing calculations
on larger clusters of size $N_c=16$.

{\em{Results.}} 
\begin{figure}[t]
\begin{center}
\includegraphics*[width=3.3in]{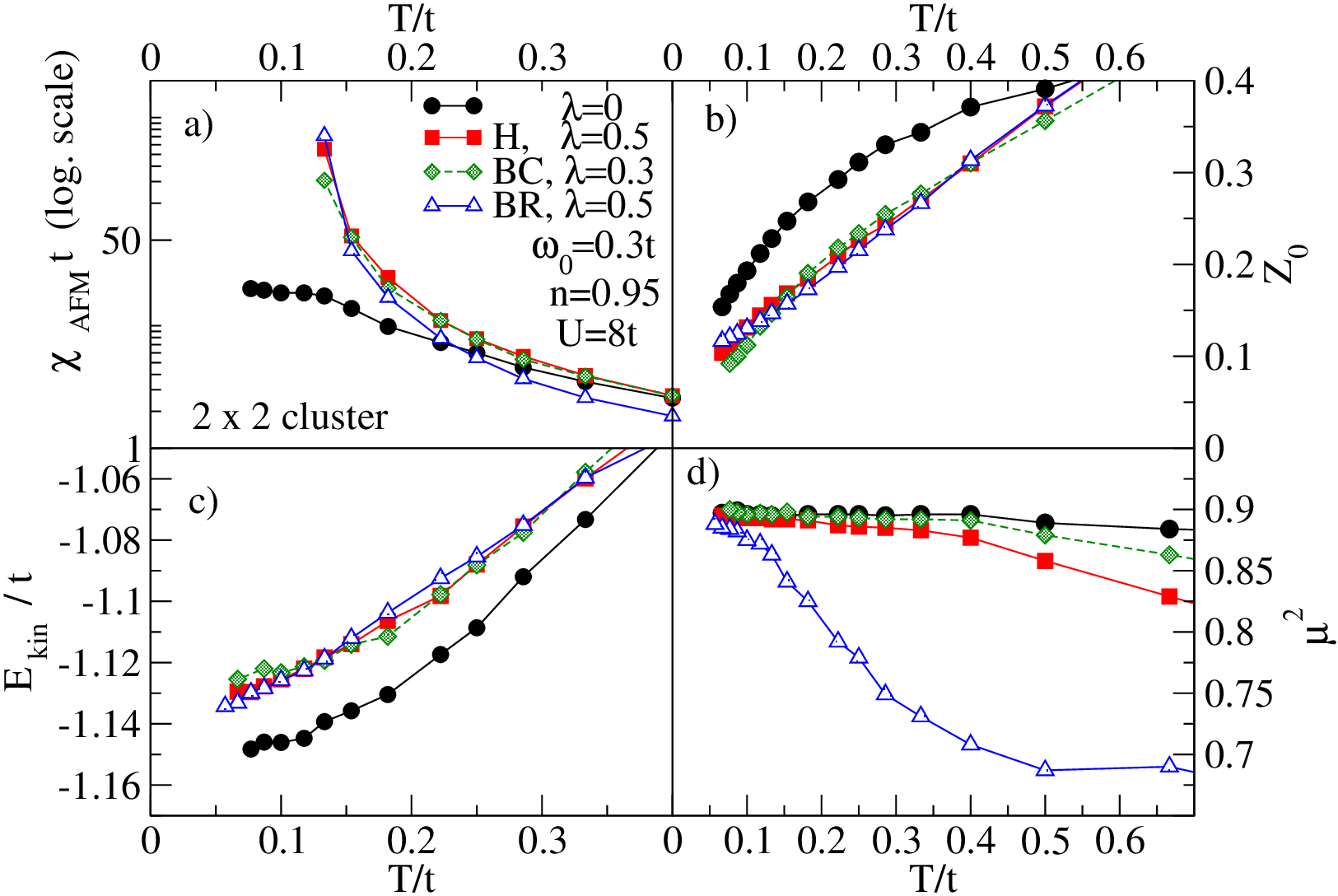}
\caption{(color online) a) AF susceptibility $\chi_{AF}$,  b) Matsubara QP fraction 
$Z_0=1/(1- {\rm{Im}} \Sigma(K,\pi T)/\pi T)$ at $K=(0,\pi)$, 
c) kinetic energy $E_{kin}=\sum_{k\sigma}\epsilon(k) n_{k,\sigma}$
and d) local moment $\mu^2=\langle(n_\uparrow-n_\downarrow)^2\rangle$
versus $T$ for HH, HBC and HBR models at $5\%$ doping. All three modes
show  enhanced polaron formation at low T.}
\label{fig:polaron}
\end{center}
\end{figure}
In a previous paper, addressing the HH model~\cite{macridin:hh},
we show that the synergistic interplay of AF correlations and the
EP interaction strongly enhances both  polaron formation and
antiferromagnetism. Since the effective EP coupling is inversely proportional 
to the kinetic energy of the holes, the AF  correlations, which reduce the 
mobility of the holes,  enhance the effective EP coupling.  On the other hand, 
since at finite doping the antiferromagnetism is suppressed by the hole motion, the  
decrease  in hole mobility due to EP coupling increases the
antiferromagnetism.  Considering the BC and the BR phonons we find a similar behavior
indicative of enhanced polaron formation.  In 
Fig.~\ref{fig:polaron} -a we show that the AF susceptibility at $5\%$ doping 
is strongly enhanced by coupling with any of the  three phonon modes.
The EP coupling also reduces the Matsubara QP weight,
$Z_0(T)=1/(1- {\rm{Im}} \Sigma(K,i\pi T)/\pi T)$
(Fig.~\ref{fig:polaron} -b for momentum $K=(0,\pi)$),  and  increases the kinetic energy 
(Fig.~\ref{fig:polaron} -c) \footnote{The reduction of $N(0)$ and the strong 
increase in the static and the dynamic charge susceptibility provide 
additional evidence of polaron formation~\cite{macridin:hh}.}. For all three 
models the local moment, $\mu^2=\langle(n_\uparrow-n_\downarrow)^2\rangle$, 
at low temperature  has a large value, close to that 
corresponding to the $\lambda=0$ case (Fig.~\ref{fig:polaron} -d), due to 
reduction of the effective hole hopping.  However, aside from the similarities 
between the three models, there are also significant differences.
We find  that for the HBC model the same value of EP coupling
produce much stronger effects then for the HH and HBR models. 
For example at $5\%$ doping a  value of $\lambda^{BC}=0.3$ produces effects 
similar to $\lambda^{H,BR}=0.5$.  The temperature dependence of the local moment (Fig.~\ref{fig:polaron} -d) as well as of other quantities such as DOS (not shown) 
is much stronger in the HBR model.

\begin{figure}[t]
\begin{center}
\includegraphics*[width=3.3in]{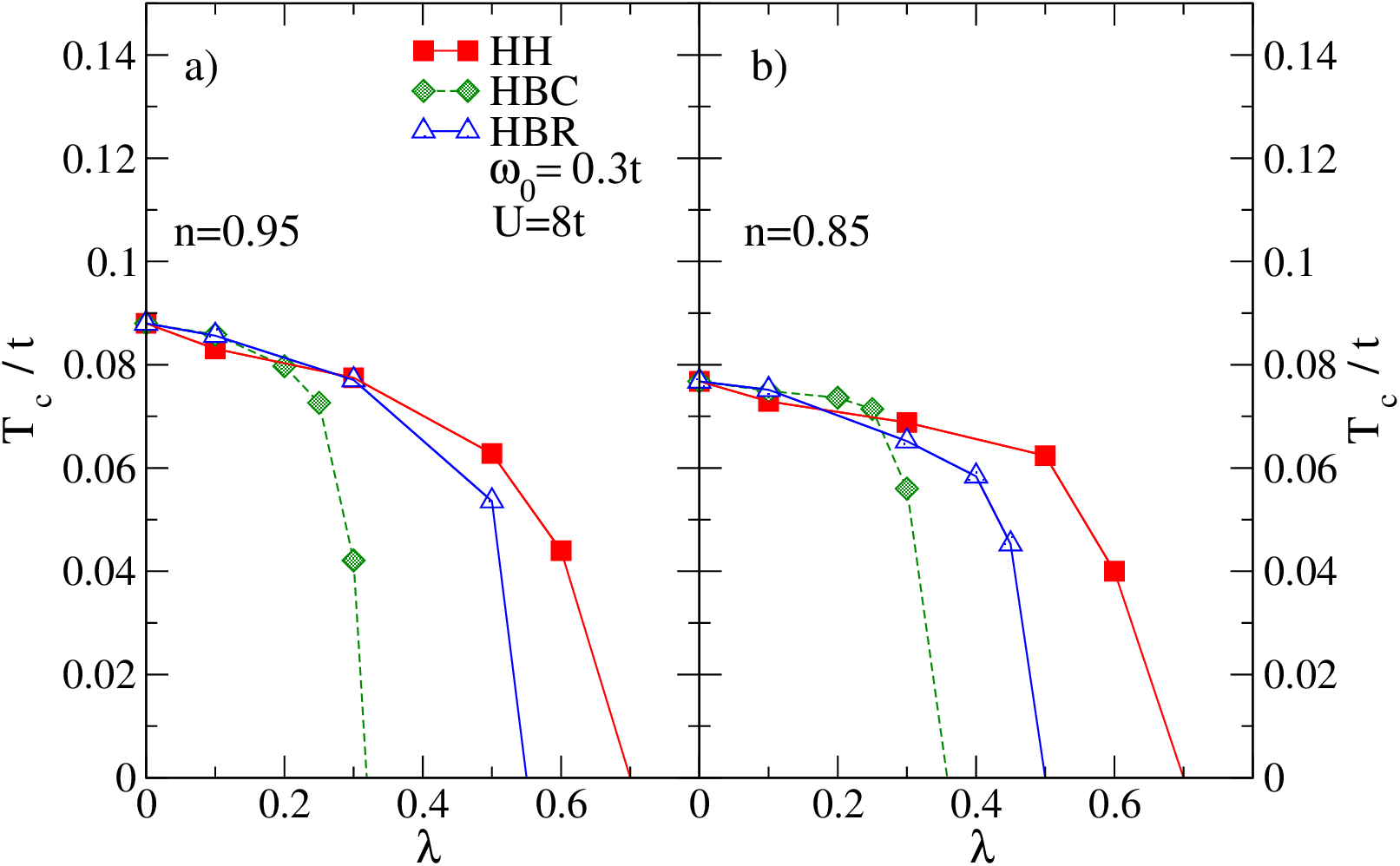}
\caption{(color online) $T_c$ versus $\lambda$ for HH, HBC and HBR models
at (a) $5\%$ and (b) $15\%$ doping for $\omega_0=0.3t$, $U=8t$ and $N_c=4$.}
\label{fig:Tcl}
\end{center}
\end{figure}

In Fig.~\ref{fig:Tcl}  we show the dependence of superconducting $T_c$ versus
EP coupling for the three models  for $\omega_0=0.3t$ at $5\%$ and $15\%$ doping. 
We find $T_c$ decreases with increasing $\lambda$,   
weakly for small $\lambda$  and then quite 
abruptly for $\lambda>\lambda_c$.  We find that
$\lambda^{H,BC,BR}_c \approx 0.5,0.25,0.4$~\footnote{$\lambda_c$ here should 
not be understood necessarily as the critical EP coupling which indicates 
the transition to the polaron regime.}. Notice that the decrease 
of $T_c$ with $\lambda$ for HBC model is much sharper once 
$\lambda>\lambda^{BC}_c$.

Previous investigations predict an enhancement of the phonon contribution to
the $d$-wave pairing interaction when the coupling with BC or H phonons is 
considered~\cite{bulut,huang:qmc}. However, an increase of the interaction 
which favors $d$-wave pairing does not necessarily imply an increase of $T_c$, 
since the reduction of the QP weight and DOS at the Fermi level, $N(0)$, 
has the opposite effect.
We find for all three models an increase of  the 
$d$-wave pairing  interaction. However, in spite of this, the $T_c$ is 
reduced as shown in Fig.~\ref{fig:Tcl}.  To understand this dichotomy, we 
must disentangle the effects of the pairing interaction from renormalization
of the single-particle propagator.

A divergent pairing susceptibility indicates the superconducting
transition and hence $T_c$.  At this instability
the leading eigenvalue of the pairing matrix becomes 1.
The pairing matrix $M=\Gamma \chi_0$ enters in the Bethe-Salpeter equation of the 
two-particle Green's function,
\begin{eqnarray}
\label{eq:bs}
\chi=\chi_0+\chi_0 \Gamma \chi=\chi_0+\chi_0(1-M)^{-1} M~.
\end{eqnarray}
where $\chi_0=G*G$ (the bubble diagram) describes propagation of the two 
particles without mutual interaction, a product
of the fully renormalized single-particle propagators 
$G(k,i \omega)=(i \omega-\epsilon(k)+\mu-\Sigma(i \omega,k))^{-1}$. 
$\Gamma_{Q=0,i\nu=0}(K,i\omega;K',i\omega')$
is the irreducible interaction vertex in the particle-particle channel
and can be regarded as the effective renormalized interaction.
We find that for all three models the leading eigenvector $\Phi_d(K,i\omega)$ 
of the pairing matrix $M$ has $d$-wave symmetry. 

\begin{figure}[t]
\begin{center}
\includegraphics*[width=3.3in]{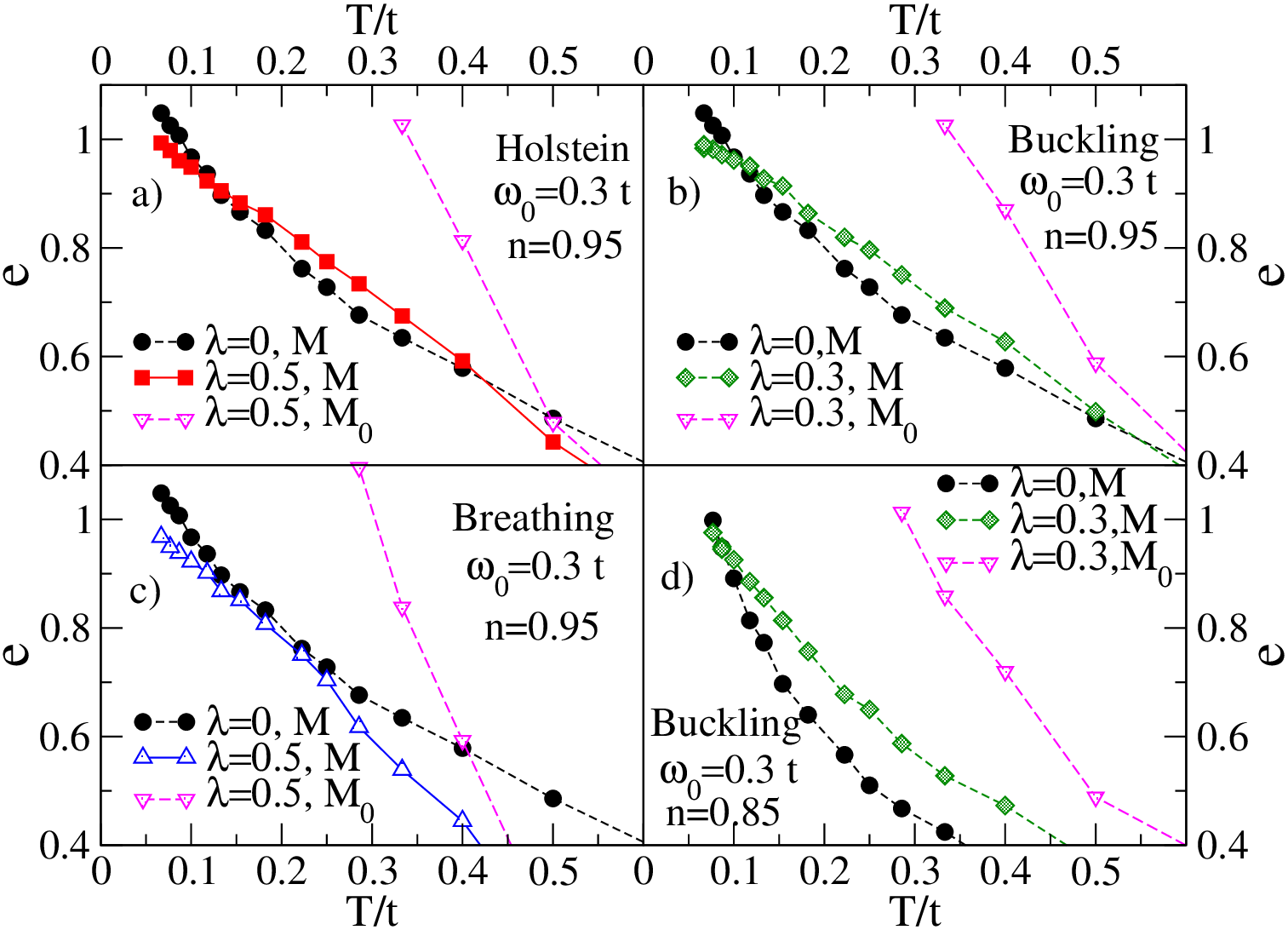}
\caption{(color online) $d$-wave eigenvalue of pairing matrices $M=\Gamma~ \chi_0$
and $M_0=\Gamma~ \chi_{00}$ versus $T$ on $N_c=4$ cluster for HH  (a), HBC (b) 
and HBR (c) models at $5\%$ doping and HBC (d) at $15\%$ doping. Despite
the overall reduction of $T_c$ shown by the eigenvalues of $M$
the eigenvalues of $M_0$ increase  strongly with decreasing $T$, showing that
EP coupling enhances the effective pairing interaction.}
\label{fig:eig}
\end{center}
\end{figure}

The phonons affect the pairing matrix by modifying both the effective 
interaction $\Gamma$ and the single particle propagator $G$, or implicitly 
$\chi_0$.  To separate these effects we compare, in Fig.\ref{fig:eig}, 
the eigenvalues of two different pairing matrices: $M$ and
$M_0$. $M$ was defined above and we define $M_0=\Gamma \chi_{00}$ with $\chi_{00}=G^0*G^0$.
$G^0(k,i\omega)=(i \omega-\epsilon(k)+\mu_U-\Sigma_U(i \omega,k))^{-1}$ 
is a single-particle propagator which does not account for the renormalization
resulting from the EP interaction, i.e $\mu_U$ and $\Sigma_U$ are the obtained 
by setting $\lambda=0$. Note  that the effective interaction
$\Gamma$ in $M_0$ is, however, fully renormalized by phonons.

The leading eigenvalue of $M$ increases with decreasing $T$ and reaches a value
of 1 at the transition temperature $T_c$.  This transition occurs at lower 
temperature when $\lambda$ is finite, thus indicating a reduction in $T_c$ in 
accordance with the phase diagram shown in Fig.~\ref{fig:Tcl}. However, the $d$-wave 
eigenvalue of the matrix $M_0$ increases much faster and reaches 1 at a  temperature 
much larger than without phonons. Since the matrix $M_0$ contains the  fully 
renormalized  interaction and the single-particle propagators which are not
renormalized by phonons, this shows that the EP coupling strongly enhances 
the $d$-wave pairing interaction. However, the competing effect, renormalization 
of $\chi_0$, is also very strong such that the net effect is a 
reduction of $T_c$. 

\begin{figure}[t]
\begin{center}
\includegraphics*[width=3.3in]{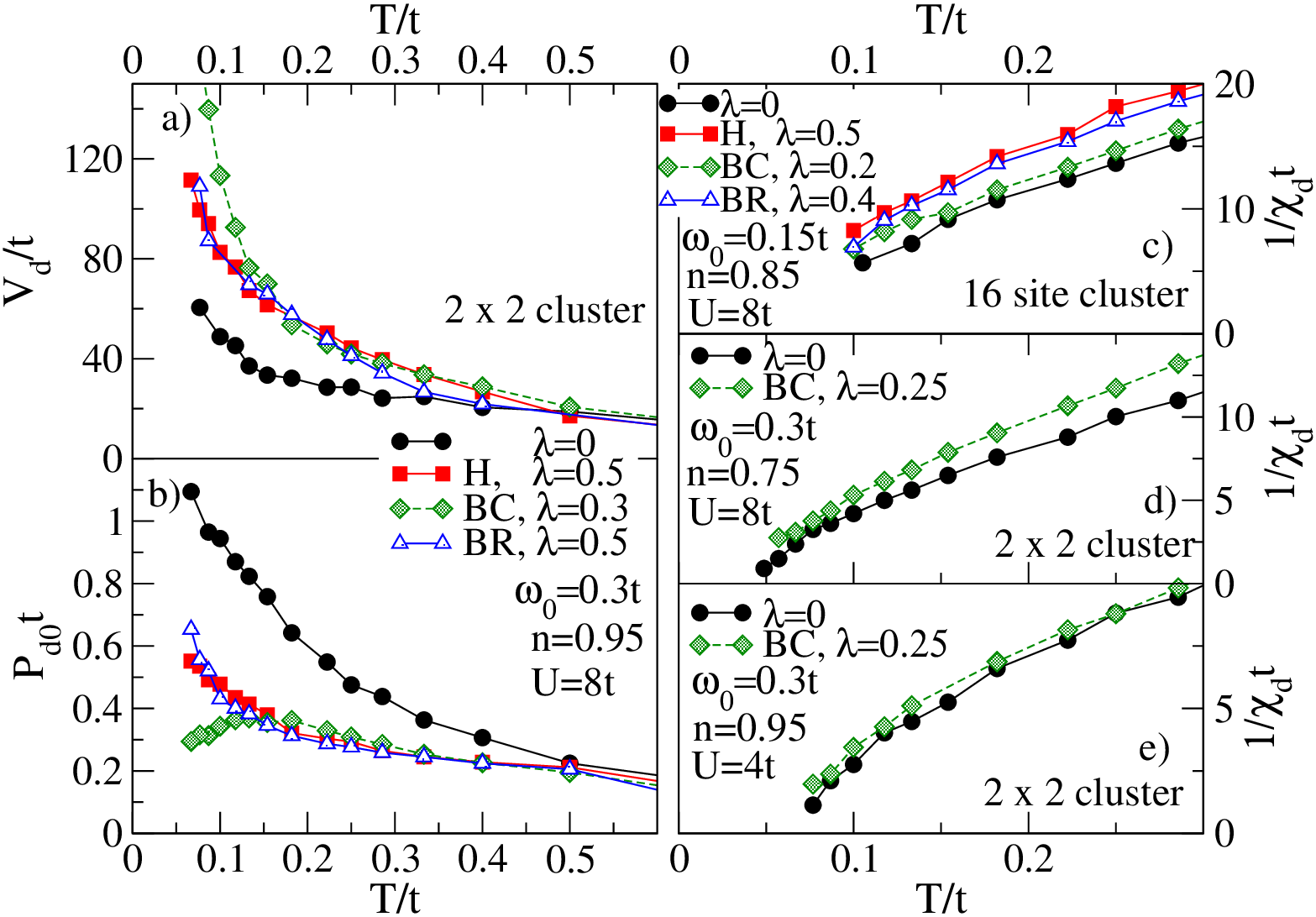}
\caption{(color online) a) $d$-wave pairing interaction $V_d$ and b) 
$d$-wave projected bare bubble $P_{d0}$
versus  T for $U=8t$,  $5\%$ doping, $\omega_0=0.3t$ and $N_c=4$ cluster. 
$V_d$ ($P_{d0}$) is enhanced (reduced) by  EP coupling.
c) Inverse of the $d$-wave pairing  susceptibility $1/\chi_d$
for HH, HBC and HBR models for $U=8t$, $N_c=16$ cluster and $\omega_0=0.15t$ at 
$15\%$ doping. d) $1/\chi_d$ for  HBC model for $U=8t$, $\omega_0=0.3t$, $N_c=4$ cluster at 
$25\%$ doping. e) $1/\chi_d$ for  HBC model for $U=4t$, $\omega_0=0.3t$, $N_c=4$ cluster 
at  $5\%$ doping. We find that a finite $\lambda$ always suppresses $T_c$.}
\label{fig:Vd}
\end{center}
\end{figure}

Alternatively, to investigate the pairing interaction
one can look at quantities such as $V_d=\sum_{K,i\omega;K',i\omega'} \Phi_d(K,i\omega) \Gamma(K,i\omega;K',i\omega')\Phi_d(K',i\omega') $ 
and $P_{d0}=\sum_{K,i\omega} \Phi^2_d(K,i\omega)\chi_0(K,i\omega)$ which 
are the respective projections of the interaction vertex and $\chi_0$
on the subspace spanned by the $d$-wave eigenvector. These were previously
defined in Ref.~\cite{maier:pairprb}. Fig.~\ref{fig:Vd} -a and -b 
show that EP coupling enhances the effective pairing interaction  $V_d$ and  
strongly reduces  $P_{d0}$, reinforcing the conclusions previously drawn from 
Fig.~\ref{fig:eig}.

Note that we see an enhancement of $d$-wave pairing interaction for all 
three phonon modes, including BR. DCA calculations in the Hubbard 
model without phonons show that the main contribution to the $d$-wave pairing 
is contained in the particle-hole  spin $S=1$ channel at 
$Q=(\pi,\pi)$~\cite{maier:pairprl}, i.e. the pairing  is a result of 
exchanging  AF spin fluctuations.   We speculate that the increase of the
pairing interaction when EP is present results from the enhancement of the
AF susceptibility.  However a decomposition of the pairing vertex 
$\Gamma$ in the fully irreducible and partially reducible particle-hole spin 
and density components, similar to the one done in Ref.~\cite{maier:pairprl} 
for the Hubbard model, is necessary to better understand the effect of phonons 
on the pairing interaction.

It is important to ask whether these results will change
for other regions of parameter space relevant for cuprates   
or when larger clusters are considered. We find that 
EP coupling always reduces $T_c$ for small phonon frequency
$\omega_0<t$. We exemplify  this by showing in Fig.\ref{fig:Vd} -c,
-d and -e the inverse of the $d$-wave pairing 
susceptibility $\chi_d$ for some particular cases. Larger cluster, $N_c=16$  
sites, and smaller phonon frequency, $\omega_0=0.15t$, results 
are illustrated in Fig.\ref{fig:Vd} -c for all
three models. Even though we were unable to reach
temperatures  equal to $T_c$, due to the sign 
problem present in large clusters calculations, it is obvious that
a finite $\lambda$ suppresses the pairing susceptibility. 
Regarding the phonon frequency dependence of $T_c$,
we find that, in general, $T_c$ decreases with decreasing $\omega_0$ (not shown),
i.e. a positive isotope effect. However the magnitude of the 
isotope effect depends strongly on the mode, $\lambda$ and doping.
A detailed investigation will be presented in a subsequent 
publication. In Fig.~\ref{fig:Vd} -d and -e we show that EP coupling 
reduces $T_c$ for the HBC model in the overdoped region, $25\%$ doping,
and in the weak coupling regime, $U=4t$.
The same conclusion can be drawn for the HH and HBR models (not shown).

Our results indicate that the coupling of the electronic density to local phonons  is, 
in general,  not favorable for superconductivity in the cuprates. 
This result is contrary to some previous speculation~\cite{gweon,capone}
about the role played by the EP interaction in high Tc,
and therefore significant to the experimental community.
Synthesis aimed at increasing $T_c$ by designing new materials with strong EP coupling may lead
to the antithetical result. A proper treatment of quasiparticle dressing is key to 
understanding the role played by the EP interaction.  Methods
that fail to properly address this point, e.g. by truncating the phonon Hilbert space, 
may lead to incorrect or erroneous conclusions.

{\em{Conclusions.}}
By employing DCA with a QMC algorithm we investigate 
the Hubbard model with H, BC and BR phonons.
We find that the interplay of EP interaction and electronic correlations in
HBC and HBR models leads to  synergistic enhancement of both polaron formation
and AF correlations, similar to previous findings
for the HH model~\cite{macridin:hh}.  Regarding superconductivity,  all three  modes produce two 
competing effects:  a strong renormalization of the single-particle electronic 
propagator which suppresses superconductivity and an enhancement of the effective pairing 
interaction which favors $d$-wave superconductivity. 
In the region of parameter space relevant for cuprates, we find that the combination of these two
effects leads to a reduction in superconducting $T_c$.

\acknowledgments 
\acknowledgments We thank T.\ Devereaux,  P. \ Kent and G.\ Sawatzky  for useful
discussions. This research
was supported by NSF DMR-0312680, CMSN DOE DE-FG02-04ER46129,
and ONR N00014-05-1-0127, and used resources provided by the
Ohio Supercomputer Center and the National Center for Computational Sciences at 
Oak Ridge National Laboratory, supported by the Office of Science of the 
U.S. Department of Energy under Contract No. DE-AC05-00OR22725.
TM acknowledges the Center for Nanophase
Materials Sciences, sponsored by the Division of Scientific User Facilities,
U.S. Department of Energy. 
BM acknowledges the University of 
North Dakota Computational Research Center, supported by NSF grants EPS-0132289 and
EPS-0447679.

\end{document}